\documentstyle[12pt,aaspp4,epsf]{article}
\def\LprimeCO{{\hbox {$L^\prime_{\rm CO}$}}}

\def \MH2{M(\Htwo)}
\def \nbarH2{\bar{n}}
\def\,{\thinspace}
\def\Msun{M$_\odot$}

\def\Lsun{L$_\odot$}

\def \kms{km\,s$^{-1}$}
\def \Kkmspc{K\,\kms\,pc$^2$}

\received{10 September 1998}
\journalid{337}{15 January 1989}
\articleid{11}{14}
\slugcomment{}
\begin{document}
\title{DETECTION OF CO(4--3), CO(9--8), AND DUST 
EMISSION IN THE BAL QUASAR APM 08279$+$5255 AT A REDSHIFT OF 3.9}
\author{D. Downes and R. Neri}
\affil{Institut de Radio Astronomie Millim\'etrique, 38406 St.\ Martin
d'H\`eres, France}
\author{T. Wiklind}
\affil{Onsala Space Observatory, S-43992 Onsala, Sweden}
\author{D. J. Wilner}
\affil{Harvard-Smithsonian Center for Astrophysics, Cambridge, MA 02138}
\and
\author{P. A. Shaver}
\affil{European Southern Observatory, D-85648 Garching bei M\"unchen, Germany}
\begin{abstract}
We detected with the IRAM interferometer the lines of CO(4--3) and
CO(9--8) from the recently-discovered broad absorption line quasar APM
08279$+$5255.  The molecular lines are at a redshift of 3.911, which
we take to be the true cosmological redshift of the quasar's host
galaxy.  This means the quasar emission lines at $z=3.87$ are
blueshifted by a kinematic component of $-2500$\,\kms , and, along with
the broad absorption lines, are probably emitted in the quasar's wind
or jet, moving toward us.  The CO line ratios suggest the molecular
gas is at a temperature of $\sim 200$\,K, at a density of 
$\sim 4000$\,cm$^{-3}$.  We also detected the dust emission at 94 and
214\,GHz (emitted wavelengths 650 and 290\,$\mu$m).  The spectral
index of the mm/submm continuum is $+3.2$, indicating the dust
emission is optically thin in this part of the spectrum.  The
extremely high CO and dust luminosities suggest magnification by
gravitational lensing.  Using the optical extent and our limit on
the size of the CO region, we estimate a magnification of 7 to 30
for the CO lines and the far-IR continuum, and 14 to 60 for the
optical/UV.  In this interpretation, the molecular gas and dust is in
a nuclear disk of radius 90 to 270\,pc around the quasar.  The quasar
is 25 to 100 times stronger than, but otherwise resembles, the nucleus of
Mrk~231. 
\end{abstract}


\newpage  
The recently-discovered broad absorption line (BAL) quasar 
APM 08279$+$5255 (Irwin et al. 1998) at $z=3.87$ has an astounding
luminosity of $5\times 10^{15}$\,\Lsun , making it the most luminous
object in the Universe.\footnote{We use
$H_0 = 50$\,\kms \,Mpc$^{-1}$ and $q_0 = 0.5$ throughout this paper.}
Ledoux et al. (1998) found that the source consists of two components
separated by 0.35$''$ with similar spectra. Together with the
extraordinary luminosity, this strongly suggests the source is gravitationally
lensed, probably by one of the  absorption-line systems along the line of sight
(Irwin et al. 1998). 
The IRAS fluxes and the submillimeter detections with the SCUBA bolometers 
indicate the presence of warm dust, with an infrared component at a temperature
of $\sim 200$~K (Lewis et al. 1998).  The source is obviously similar to the
gravitationally lensed sources IRAS F10214$+$4724 at $z=2.3$ and the Cloverleaf
quasar at $z=2.6$ (e.g., Serjeant et al. 1998; Kneib et al. 1998).
   In this Letter, we
present observations of APM08279$+$5255 in the 1 and 3\,mm continuum  
and in CO lines from warm, dense, molecular gas.

We observed this remarkable object with the IRAM Plateau de Bure interferometer
and detected both the CO(4--3) and CO(9--8) lines at $+2500$\,\kms\ relative to
$z=3.87$. We also detected dust continuum emission, redshifted to 1.4 and
3.2\,mm. The observations were done in August and November 1998 with the
antennas on baselines from 24 to 160\,m, giving synthesized beams of
6.6$'' \times 5.3''$ at 94\,GHz and 3.2$'' \times 2.3''$ at 211\,GHz.   Receiver
temperatures were 45 to 65\,K at both frequencies. The spectral correlators
covered 1400\,\kms\ at 3.2\,mm and 760\,\kms\ at 1.4\,mm, with resolutions of 8
and 4\,\kms\ respectively. Amplitudes were calibrated with the strong sources
3C454.3, 3C273, 0923+392, and MWC 349,  and phases were calibrated with 4C54.15
and 0804$+$499. {\bf Table~1} summarizes our results.

We interpret the redshift, $z=3.911$, of the CO(4--3) and CO(9--8) lines as
being the true cosmological redshift of the quasar's host object.  
The optical redshift of $z=3.87$, derived from high excitation lines 
(Irwin et al. 1998), thus corresponds to gas flowing toward us, 
with a kinematic blueshift of 2500\,\kms\ relative to the molecular gas.
Quasar high-excitation lines  are usually blueshifted by 500 to 1000\,\kms \
relative to low excitation lines (e.g., Storrie-Lombardi et al. 1996).
In addition to the main optical emission lines,
 Ledoux et al. (1998) found an NV doublet in absorption at $z=3.901$
and Irwin (1998) found several CIV doublets up to $z=3.92$, so there 
 is also some highly ionized gas at velocities close to the CO redshift.
The CO(4--3) line from APM 08279$+$5255 has a line width of 480\,\kms\ (FWHM)
(Fig.~1), typical of line widths from rotating nuclear disks 
of molecular gas in ultraluminous IR galaxies (e.g. Downes \& Solomon 1998).   

The images in Fig.~1 show a source that coincides within 0$''.3$ of the optical
quasar (revised optical position  from Irwin 1998),  and with the non-thermal
radio source detected at 1.4 GHz (White et al. 1997). The interferometer beams
are thus far too large to show source sizes or velocity gradients.
Higher-resolution observations are now in course. A tentative size of the CO
emission region has been obtained from $u,v$ plane fits to the CO(9--8) data.
Although noisy, it implies a  CO source size of 0$''.8\pm 0''.2$.

The continuum fluxes were derived from the line-free channels at 3.2\,mm, and
from the upper sideband measurements at 1.4\,mm to avoid contamination from the
CO lines.  The corresponding emitted wavelengths are 290 and 650\,$\mu$m, i.e.,
dust radiation in the far-IR/sub mm range. The fluxes are given in Table~1. In
this part of the spectrum,  the continuum spectral index is $+3.2$,
corresponding to optically thin dust. Our 1.4\,mm flux of 17\,mJy is somewhat
lower than that quoted by  Lewis et al. (1998) from wideband bolometer
measurements, but probably consistent within the errors.

Although we lack a more precise size measurement for the time being,
we can deduce a number of constraints on the source from
the CO and dust detections.

1) {\it The gas is hot}.  
The CO $J=9$ level is $J(J+1)\times 2.77$\,K = 249\,K above the ground
state, and in normal galactic spiral-arm molecular clouds, the 
typical H$_2$ densities of $\sim 300$\,cm$^{-3}$ are not sufficient to
collisionally populate the CO $J=9$ level. Therefore the mere detection
of the CO(9--8) line shows the gas is hot and dense. Could the gas be
radiatively excited by the quasar?  Maloney, Begelman, \& Rees (1994) 
showed that dense gas can survive in molecular form close to an AGN.
The strong radial dependence of radiative excitation in their model,
however, plus the large dust mass, the large dust opacity at short
wavelengths, and the inferred extent of the molecular gas distribution
(see below), make it unlikely that the quasar's  radiation can affect
a significant volume of gas. We therefore think the CO is collisionally
excited.

Lewis et al. (1998) fit a blackbody curve to the sub-mm continuum
and obtained a dust temperature of 220~K.  One can also fit a two-temperature
model, with a cool, optically thin, far-IR component and a hot, opaque,
mid-IR component.  However, even with this fit, the ``cool'' component
must be at $> 150$\,K.
Our escape-probability radiative transfer models show that if the
gas temperature is 140 to 250\,K, with an H$_2$ density of 
$\sim 4000$\,cm$^{-3}$, in a cosmic blackbody background field 
of $(1+z)\times 2.7$\,K, then collisional excitation 
yields CO level populations that cause the line 
brightness temperature of CO(9--8) to be 
about half that of CO(4--3). Hence the integrated flux, in Jy\,\kms , of
CO(9--8) should be about 2.5 times that of CO(4--3), as is observed.

2) {\it If the CO source size is 0.$''$6 to 1.$''$0, 
as indicated by our data thus far,
 then the gravitational lens must magnify the CO lines by a 
factor of 7 to 20}.
The reasoning is as follows.  The apparent CO luminosity is
\begin{equation}
\LprimeCO ({\rm obs}) \ \ = \ \  363\,  
(S\Delta V)\, \lambda^2\,  D^2_A\, (1+z)\ \ \ ,
\end{equation}
where $(S\Delta V)$ is the integrated line flux in Jy\,\kms , $\lambda$ is
the observed wavelength in mm, and $D_A$ is the emission distance 
(= angular size distance) in Mpc (this may be derived from eq.(1) of Solomon,
Downes, \& Radford (1992)). The lens magnification of the CO can then be estimated
from eq.(2) of Downes, Solomon, \& Radford (1995) to be
\begin{equation}
m_{\rm CO} \ \ = \ \ {a\over R}\ \ 
= \ \ {{\pi\, a^2\,\Delta V}\over \LprimeCO ({\rm obs})} 
\, f_V \,T_b \ \ \ \ .
\end{equation}
where $a$ is the apparent semi-major axis of the magnified CO
source, in pc, $\Delta V$ is the line width in \kms , $\LprimeCO$
is the apparent (magnified) CO luminosity in \Kkmspc ,  $R$ is
the true source radius in pc, $f_V$ is the velocity filling factor,
and $T_b$ is its rest frame  brightness temperature, which can be 
estimated from the observed CO(9--8)/CO(4--3) ratio of 0.5 in 
CO luminosity. 
If both CO lines arise in the same volume, then our escape probability
models yield intrinsic CO(4--3) and (9--8) brightness temperatures of 135
and 70\,K respectively.
For the observed CO luminosity and line width (Table~1), and $f_V = 1$, 
the formula indicates the lens magnifies the CO source 7 to 20 times.  Hence
the true radius of the CO source is 160 to 270\,pc.

3) {\it If the dust continuum source size is 0.$''$4 to  0.$''$8, 
then the far-IR/sub-mm
magnification is 7 to 30, 
and the true far-IR luminosity is (3 to 14)$\times 10^{13}$\,\Lsun .}
Because of opacity effects, the far-IR source is likely to be 
smaller than the CO source, but larger than the optical source. 
Although we cannot measure a diameter for the dust source at the 
peak of the far-IR continuum, a reasonable guess is that it is about the 
same size as the CO source, or 
about twice the $\sim 0''.35$ extension derived from
the H-band image by Ledoux et al. (1998).
With a dust continuum source size of 0.$''$4 to 0.$''$8, 
 and the apparent far-IR luminosity of 
$1\times 10^{15}$\,\Lsun \ from the observations of
Lewis et al. (1998), we use
 eq.(6) of Downes et al.\ (1995) to calculate the lens 
magnification of a 220\,K blackbody, far-IR source to be  
\begin{equation}
m_{\rm fir} \ \ \equiv \ \ {a_{\rm fir} \over R_{\rm bb} } \ \ 
= \ \ 4.17\times 10^{-6}  \ 
{ 
{
a^2_{\rm fir} \ (T_d/220\,{\rm K})^4
} 
\over 
{
L_{\rm fir}({\rm app})/(1\times 10^{15} )
} 
}  
\end{equation}
where $a_{\rm fir}$ is the apparent semi-major axis of the magnified far IR
source, in pc, $T_d$ is the dust temperature in K, and $L_{\rm fir}$ 
is the apparent far-IR luminosity \ in \Lsun . This then yields a far-IR/sub-mm 
magnification of 7 to 30, meaning that the true radius of the far-IR/sub-mm
source is 90 to 180\,pc, and the true far-IR luminosity is 
(3 to 14)$\times 10^{13}$\,\Lsun .

4) {\it If the dust absorbs half the power of the quasar, then the optical/UV
magnification factor is 14 to 60.} 
In most of the ``warm" ultraluminous galaxies, the obscuring dust ring or disk
absorbs about half of the available power from the central source,
and the rest escapes out the poles of the torus or disk (see the
analysis of Mrk~231 in Downes \& Solomon 1998).  This implies the
true optical/UV power input is twice the 
far-IR luminosity, so the apparent optical/UV luminosity of 
$4\times 10^{15}$\,\Lsun\  corresponds to an optical/UV magnification
of 14 to 60 by the gravitational lens. This is 
comparable with the optical magnification deduced for IRAS F10214$+$4724 
(e.g., Broadhurst \& Lehar 1995; Eisenhardt et al. 1996).

5) {\it The dust mass is (1 to 7)$\times 10^7$\,\Msun\ .}
Because the dust is optically thin 
at observed wavelengths $>1.4$\,mm (emitted wavelengths $>290$\,$\mu$m), 
we can use the observed continuum flux and the derived magnification
factor $m_{\rm fir}$ to estimate the dust mass from
\begin{equation}
 M_{\rm d} = 
{
{
S(\nu_{\rm obs}) D_A^2 (1+z)^3
}
\over 
{
m_{\rm fir}\,\kappa_{\rm d}(\nu_{\rm e})\, B(\nu_{\rm e},\, T_{\rm d})
}
}
\ \ \ \ \ ,
\end{equation}
where $S$ is the flux density, $\nu_{\rm obs}$ and $\nu_{\rm e}$ are the
observed and emitted frequencies, $D_{\rm A}$ is the emission distance
(angular size distance), $B$
is the Planck function, and $T_{\rm d}$ is the dust temperature
 (see, e.g., Downes et al. 1992).
We use the dust absorption coefficient  
$\kappa_{\rm d} = 0.4\,(\nu_{\rm e}/250\, {\rm GHz})^2$~cm$^2$ per gram
of {\it dust}  (e.g. Kr\"ugel \& Siebenmorgen 1994). 
This yields a dust mass of (1 to 7)$\times 10^7$\,\Msun .
For further discussion of uncertainties in the dust mass, see
Hughes et al. (1997).

6) {\it The gas mass is (1 to 6)$\times 10^9$\,\Msun .}
The gas is more problematic because the high optical depths of
the CO(4--3) and (9--8) lines ($\tau =$ 20 to 40 in our models) 
would be consistent with the magnification model derived above and
the apparent CO luminosities for a range of gas masses. Our estimate
is based on a mean H$_2$ 
density of 4000\,cm$^{-3}$ from the excitation modeling,
and a true radius of 160 to 270\,pc for the CO source, which we 
assume to be a disk with thickness-to-radius ratio of 0.3.
The gas mass we quote is for H$_2$ plus helium.
In the excitation modeling, we assumed
 typical interstellar abundances in our
Galaxy.  The gas mass 
could be lower if the metal abundances are super-solar, 
as in the near-nuclear regions of many quasars. 

The derived values also depend on the assumed cosmology.
 For $q_0 = 0.5$ and $h=H_0/50$, our estimates of
true size, gas mass, and dynamical mass scale as $h^{-1}$,  $h^{-2}$, and
$h^{-1}$, respectively.   Note that at $z=3.9$, 
quantities derived with $H_0=50$\,\kms\,Mpc$^{-1}$ and $q_0=0.5$ 
are the same as those derived with 
$H_0 = 75$\,\kms\,Mpc$^{-1}$ and $q_0 = 0.15$.
We summarize our results in Table~2, where we interpret the
optical/UV continuum as the emission of an accretion disk, and the 
CO lines and far-IR continuum 
as the emission of a larger, circumnuclear disk.  
For most entries in the Table,
we give a range of values corresponding to the uncertainties in the source
size, which yield solutions with 
low magnification (first value) and high magnification
(second value).  The low magnification solutions 
make the gas mass equal to the dynamical mass, so the true magnification 
is more likely to lie in the middle to high end of the range.

In summary,
as with other gravitationally lensed sources detected in CO and the 
millimeter dust continuum, we have a good idea of the intrinsic CO and far
IR surface brightness of APM 08279$+$5255 at $z = 3.9$.
We can therefore estimate the CO and far IR/sub-mm magnifications.
Our results (Table~2) suggest the far IR and CO sources are
magnified 7 to 30 times and have true radii of 90 to 270\,pc.
The true far IR luminosity is (3 to 14)$\times 10^{13}$\,\Lsun .
Because the CO and far IR magnifications are lower than the optical/UV 
magnification, the true total luminosity is of the order of
(0.7 to 3)$\times 10^{14}$\,\Lsun , with about half of the quasar's power 
being absorbed in the surrounding dust. From the 
CO line width, we think this far-IR emitting dust 
is in a rotating nuclear disk.  In dust and molecular gas content, this
disk resembles closely the circumnuclear disk in the IR ultraluminous Seyfert~1
galaxy Mrk~231, with the critical difference that 
 APM 08279$+$5255 is 25 to 100 times stronger than Mrk~231 in 
total intrinsic luminosity.

\acknowledgments  We thank the referee for numerous helpful comments.
T.W. thanks the STScI Visitor Exchange Program for
their hospitality during this work, and the Swedish National Science
Council for their support.  


\clearpage

\begin{deluxetable}{lcccc}
\tablewidth{0pc}
\tablecaption{OBSERVATIONS OF APM 08279+5255}
\label{Observations}
\tablehead{
\colhead{Parameter} &\colhead{CO(4--3)} &\colhead{CO(9--8)} &\colhead{3.2\,mm dust} 
&\colhead{1.4\,mm dust}
}
\startdata
Emitted frequency (GHz)   &461.0408	&1036.912	         & 463.1	&1051.0
\nl
Observed frequency (GHz) &$93.872 \pm 0.005$	&$211.145\pm 0.007$  &94.3	&214.0
\nl
Redshift  (LSR)  &$3.9114 \pm 0.0003$	&$3.9109 \pm 0.0002$	& ---	& --- 	
\nl 
Line width (\kms )   &$480\pm 35$ &480 (adopted)	&---	&---
\nl
 Peak flux density, $S$ (mJy)	&$7.4 \pm 1.0$  &$17.9\pm 1.4$ &$1.2\pm 0.3$ &$17.0\pm 0.5$
\nl
Integrated flux (Jy\,\kms) &$3.7\pm 0.5$  &$9.1\pm 0.8$  &---	&---
\tablenotetext{}{The table gives the statistical errors. In addition, 
systematic uncertainties in the flux scales are $\pm 5$\%
at 3.2\,mm and $\pm 15$\% at 1.4\,mm.}
\enddata
\end{deluxetable}
\clearpage

\begin{planotable}{l ccc}
\tablecaption{PROPOSED COMPONENTS OF APM 08279$+$5255}
\label{Proposed components of 10214+4724}
\tablehead{\colhead{}  &\colhead{CO(4--3)}&\colhead{Far IR}  
&\colhead{optical/UV}
\nl\colhead{}
&\colhead{line}
&\colhead{continuum}
&\colhead{continuum}
}
\startdata
Component		&nuclear disk 	&nuclear disk 	&accretion disk 
\nl
Apparent luminosity	&$9.5\times 10^{10}$\,L$_\ell$
			&$1\times 10^{15}$\,\Lsun  
			&$4\times 10^{15}$\,\Lsun 
\nl
Apparent diam. (arcsec)	&0.6 --- 1.0	
			&0.4 --- 0.8  
			&(7 --- 14) $\times 10^{-5}$
\nl
Magnification		&7 --- 20		
			&7 --- 30	
			&14 --- 60
\nl
True luminosity		&(17 --- 6)$\times 10^9$\,L$_\ell$
			&(14 --- 3)$\times 10^{13}$\,\Lsun  
			&(29 --- 7)$\times 10^{13}$\,\Lsun 
\nl
Temperature (K)		&$T_{\rm kin}=200$  
			&$T_{\rm dust}=220$ 
			&$T_{\rm acc.disk}=30000$
\nl
True radius (pc) 	&270 --- 160	&180 --- 90	&0.02 --- 0.01
\nl
Gas mass (H$_2$ + He) (\Msun )
			&(6 --- 1)$\times10^{9}$ & ---& ---
\nl
Dust mass (\Msun )	&---		&(7 --- 1)$\times 10^7$	& ---
\nl
Dynamical mass (\Msun )	&(6 --- 3)$\times10^{9}$  & ---& ---
\nl
(if $V_{\rm rot}=300$\,\kms )
\tablenotetext{}
{ Entries with a range of values 
correspond to low magnification (first value) and high magnification 
(second value). Masses are enclosed masses interior to the true radius.
The CO line luminosity unit L$_\ell$ is in \Kkmspc . } 

\tablenotetext{}{Adopted parameters for luminosities:  
$H_0 = 50$\,\kms \,Mpc$^{-1}$, $\Omega = 1$, $\Lambda = 0$, 
emission distance 
(angular size distance) $= D_A = 1.35$\,Gpc \ (1$'' = 6.54$\, kpc),
reception distance (metric distance) $= D_M = 6.63$\,Gpc, 
age = 1.2\,Gyr, lookback time
= 11.9\, Gyr, recession velocity at emission $= 2.5 c$ (e.g. Harrison, 1981).}

\enddata
\end{planotable}
\clearpage


\begin{figure}
\vspace{-1.5cm}
\plotfiddle{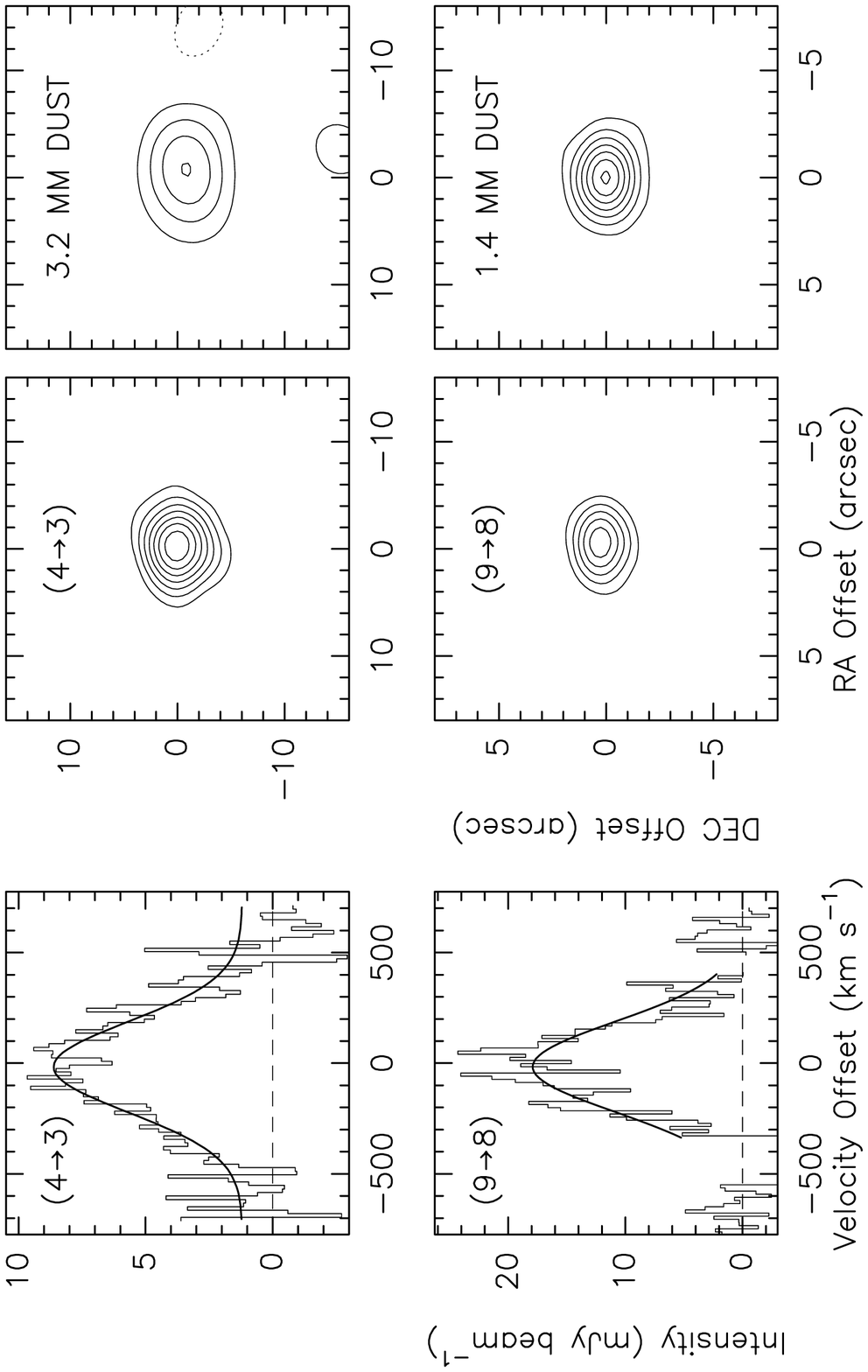}{\hsize}{-90}{73.5}{73.5}{-340}{520}
\end{figure}
\begin{figure}
\vspace{-6.5cm}
\caption[{\bf CO  spectra and maps.}]
{
CO spectra and maps of APM 08279+5255.
%
{\it Upper left:} CO(4--3) spectrum with resolution  16\,\kms , 
velocities relative to 93.867\,GHz; the 1.2\,mJy continuum was not  
subtracted.  The central part of the line has more integration time
and hence less noise than the line wings.
%
{\it Lower left:} CO(9--8) spectrum with resolution  14\,\kms ,  
velocities relative to 211.134\,GHz, and the 17\,mJy continuum subtracted.
The zero-level baseline insets right and left show the noise in the 
line-free sideband after continuum subtraction.  
Fitted Gaussians to both spectra have a width of 480\,\kms \ (FWHM).  
%
{\it Upper center:}
CO(4--3)  intensity in a 1000\,\kms\ wide band centered on 93.867 GHz,
with the 1.2\,mJy subtracted.
Contour step = 0.5\,mJy beam$^{-1}$.
%
{\it Upper right:} 
The 3.2\,mm dust continuum in a 500\,MHz wide band centered on 94.3\,GHz.
  Contour step = 0.5\,mJy beam$^{-1}$ (1.7 $\sigma $). 
%
{\it Lower center:}
CO(9--8) intensity in a 760\,\kms\ wide band, with the 17\,mJy continuum 
subtracted. Contour step = 2\,mJy beam$^{-1}$.
%
{\it Lower right:}
The 1.4\,mm dust emission in a 450-MHz band centered on 
214.0\,GHz.
Contour step = 2\,mJy beam$^{-1} (4\sigma$).  
%
Beams are $6''.6\times 5''.3$ (at 3.2\,mm)
and $3''.2 \times 2''.3$ (at 1.4\,mm).
Offsets in all maps are relative to the CO and mm-dust source position, 
  08$^{\rm h}31^{\rm m}41^{\rm s}.70$,  \ 52$^\circ 45'17''.35$
(J2000; $\pm 0''.3$).
}
\end{figure}

\clearpage

\end{document}